\documentstyle[12pt]{article}
\newcommand{\draft}{
        \renewcommand{\baselinestretch}{1.0}%
        \small\normalsize%
}

\draft

\begin{document}
\title{\bf Early Phases of Different Types of Isolated Neutron Star}
\author{A\c{s}k\i n Ankay$\sp{1,2}$
\thanks{email: askin@gursey.gov.tr},
Efe Yazgan$\sp3$
\thanks{email: efey@newton.physics.metu.edu.tr}, \\
Serkan \c{S}ahin$\sp1$,
G\"{o}k\c{c}e Karanfil$\sp1$ 
\\ \\
{$\sp1$ Department of Physics, Bo\u{g}azi\c{c}i University,} \\
{\.{I}stanbul, Turkey} \\
{$\sp2$ T\"{U}B\.{I}TAK Feza G\"{u}rsey Institute,} \\ 
{\.{I}stanbul, Turkey} \\
{$\sp3$ Department of Physics, Middle East Technical University,} \\
{Ankara, Turkey}
}

\date{}
\maketitle

\begin{abstract}
\noindent
Two Galactic isolated strong X-ray pulsars seem to be in the densest
environments compared to other types of Galactic pulsar. X-ray
pulsar J1846-0258 can be in an early phase of anomalous X-ray
pulsars and soft gamma repeaters if its average braking index is
$\sim$1.8-2.0. X-ray pulsar J1811-1925 must have a very large
average braking index ($\sim$11) if this pulsar was formed by SN
386AD. This X-ray pulsar can be in an early phase of the evolution
of the radio pulsars located in the region P$\sim$50-150 ms and
\.{P}$\sim$10$^{-14}$-10$^{-16}$ s/s of the P-\.{P} diagram.
X-ray/radio pulsar J0540-69 seems to be evolving in the direction to
the dim isolated thermal neutron star region on the P-\.{P} diagram.
Possible progenitors of different types of neutron star are also
discussed.

\end{abstract}

\section{Introduction}

There are several different types of isolated neutron star: radio
pulsars some of which also radiate in other bands of the
electromagnetic spectrum; dim radio quiet neutron stars (DRQNSs)
which are dim in X-rays, not detected in radio band and all of which
are connected to supernova remnants (SNRs); dim isolated thermal
neutron stars (DITNSs) which are also dim in X-rays and not detected
in the radio, and which have long spin period (P) contrary to DRQNSs
(except Geminga which has short spin period but which shows DITNS
characteristics); anomalous X-ray pulsars (AXPs) and soft gamma
repeaters (SGRs) which have L$_X$/\.{E}$>$1 (where L$_X$ is the
X-ray luminosity and \.{E} is the rate of rotational energy loss)
with long spin periods and which show gamma ray bursts and X-ray
flares.

Radio pulsars are believed to be born on the upper-left part of the
P-\.{P} diagram and evolve to the right part. Some radio pulsars
(some of which have also been detected in X-rays) are located on the
lower-left part of the diagram (see Fig.1). The birth locations of
these pulsars on the diagram is an open question. More importantly,
the birth locations of SGRs/AXPs and DITNSs on the diagram are also
unclear. If we can identify some sources which are younger
appearances of these different types of neutron star, we can guess
where they come from on the diagram, and furthermore, we can have
more information on different characteristics of these neutron
stars.

There are 2 Galactic isolated pulsars which have some different
characteristics compared to other isolated pulsars. These 2 pulsars
are J1846-0258 (P=0.32359795 s, \.{P}=7.09706$\times$10$^{-12}$
ss$^{-1}$, which is connected to SNR Kes 75 $^{1,2}$) and J1811-1925 
(P=0.064667 s, \.{P}=4.40$\times$10$^{-14}$ ss$^{-1}$, which is connected 
to SNR G11.2-0.3 $^{1,2}$). Both of these pulsars have been
identified as X-ray pulsars with strong X-ray emission compared to
DRQNSs and DITNSs, but none of them has been detected in radio band.
They seem to have very dense environments $^3$. In
addition to these 2 X-ray pulsars, there is a pulsar in Large
Magellanic Cloud (LMC) which has strong X-ray emission but which is
very dim in radio band. This pulsar (J0540-69) also seems to have a very 
dense ambient medium.

In Section 2, the observational data of these pulsars and their SNRs
are examined and the possibility that these pulsars can be in early
phases of different types of neutron star is discussed. In Section
3, conclusions are given.

\section{Three isolated strong X-ray pulsars in very dense media}
\subsection{X-ray pulsar J1846-0258; early phase of SGR/AXP}
The position of J1846-0258 on the P-\.{P} diagram
together with its possibly small average braking index
(n=1.86$\pm$0.08 with the assumption of no glitch and/or significant
timing noise $^4$) suggest that this pulsar may be in a
phase preceding SGRs/AXPs (Fig.1). The X-ray luminosity of J1846-0258 is on
the order of 10$^{35}$ erg/s $^{5,6}$ which is typical
for SGRs/AXPs. Moreover, the L$_X$/\.{E} value is
$\sim$0.016 $^{2,6}$ which is very large compared to
radio pulsars (more than 6 times larger than L$_X$/\.{E} value of
Crab pulsar). For SGRs/AXPs L$_X$/\.{E} $>$ 1 (the name 'anomalous'
comes from this fact) and if the X-ray luminosity of J1846-0258 does
not drop significantly in the next $\sim$10$^4$ yr its L$_X$/\.{E}
value will be similar to those of SGRs/AXPs.

Even if J1846-0258 has n=2-2.5 (n=2 corresponds to constant
\.{P}) because of some glitches, this pulsar comes to a position
above the position of AXP 2259+586 and probably above the positions
of DITNSs assuming that the three DITNSs for which only the upper
limits on \.{P} are known have positions close to the position of
J0720.4-3125 on the P-\.{P} diagram (Fig.1).

According to Blanton \& Helfand $^7$ SNR Kes 75 (G29.7-0.3) has an age
$\sim$10$^3$ yr. It is better to adapt the age of the pulsar (which
can be $\sim$1700 yr) as the age of Kes 75 -- J1846-0258 pair
considering the possibly small value of n of the pulsar and also the
large diameter of the SNR $^6$.
For n=2 (i.e. for constant \.{P}), J1846-0258 can reach to a spin
period $\sim$2.5 s in 10$^4$ yr. Ages of the SNRs connected to
AXPs are a few times 10$^4$ yr $^3$. In
(2-5)$\times$10$^4$ yr J1846-0258 can come to the region of
AXPs/SGRs on the P-\.{P} diagram.

J1846-0258 seems to be in a very dense medium $^3$ and SNR
Kes 75 is expanding into a higher density medium on its far side
$^8$. As suggested by Guseinov et al. $^3$ the progenitor of
J1846-0258 can be an O-type star.

In Guseinov et al. $^{9,10}$ the idea of possible existence
of low-mass neutron stars has been introduced to explain the
positions and possible evolution of SGRs/AXPs on the P-\.{P} diagram
without any need for very high magnetic field (B$\le$10$^{14}$ G). 
Magnetic dipole radiation of neutron stars is given as $^{11}$
\begin{equation}
L = \frac{2}{3} \frac{\mu^2 \omega^4}{c^3}Sin^2\beta = \frac{2}{3}
\frac{B_r^2 R^6 \omega^4}{c^3}Sin^2\beta
\end{equation}
where $\mu$ is the magnetic moment, $\omega$ the angular velocity, c the
speed of light, $\beta$ the angle between the rotation axis and
the magnetic field axis, B$_r$ the real dipole magnetic field and R  
the radius of neutron star. On the other hand, the rate of rotational  
energy loss of spherically symmetric neutron stars which have rigid body
rotation is given as
\begin{equation}
\dot{E}=\frac{4\pi ^2 I \dot{P}}{P^3}
\end{equation}
where I is the moment of inertia, P the spin period of pulsar, and 
\.{P} the time derivative of P. 
From expressions (1) and (2) we get
\begin{equation}
\dot{P} \propto \frac{B_r^2 R^4}{M P}
\end{equation}
where M is the mass of pulsar. So, a neutron star with
M$\sim$0.5-0.7 M$_{\odot}$ must have $\sim$4-9 times larger \.{P} 
compared to a neutron star with M$\sim$1.4-1.5 M$_{\odot}$ if both
neutron stars have the same B$_r$ and P values 
(considering also the increase in R when M is smaller).
For such a low mass neutron star the propeller mechanism can work
more efficiently to spin down the pulsar in relatively very short
time and the reconnection of the magnetic field can occur more
easily to produce the $\gamma$-ray bursts.

X-ray pulsar J1846-0258 can be such a low mass pulsar. Its possibly small 
n and its position on the P-\.{P} diagram support the low-mass neutron
star idea. Moreover, the X-ray luminosity and the present
L$_X$/\.{E} value of this X-ray pulsar further suggest that it can
be in a phase preceding SGR/AXP phase. Also, J1846-0258 is an
isolated X-ray pulsar without any detected radio emission similar to
SGRs/AXPs. Besides, such low mass neutron stars must have B$_{initial}$ 
= 3$\times$10$^{13}$ - 10$^{14}$ G $^{9,10}$; J1846-0258 is 
only $\sim$1700 yr old and its B=5$\times$10$^{13}$ G. If J1846-0258 is 
the former appearance of all SGRs/AXPs, then we can say that such objects 
must have O-type progenitors.

In Guseinov et al. $^{12}$ SGRs/AXPs were claimed to be in active stages 
of their evolution and they predicted that these sources would spend
some time in passive stages (i.e. they must become unobservable
occasionally). XTE J1810-197 has recently been identified as a
transient AXP and it was seen that this source was unobservable in
the past $^{13}$ proving the prediction of Guseinov et al.

So, this type of neutron star is possibly born with P$_0$$\sim$10-30
ms and evolves to the right part of the P-\.{P} diagram either with
increasing \.{P} or with \.{P}$\sim$const. as an isolated strong X-ray 
pulsar. It reaches to
the SGR/AXP region in (2-5)$\times$10$^4$ yr and shows $\gamma$-ray
bursts and X-ray flares. Occasionally, it drops to a passive state
and its luminosity drops significantly that it becomes unobservable.

The lack of any detected radio emission from X-ray pulsar J1846-0258
and from SGRs/AXPs can be due to the existence of ionized gaseous
wind which must suppress the radio pulsar phenomenon when their
periods are very small (in less than $\sim$10$^3$ yr).

\subsection{X-ray/radio pulsar J0540-69; early phase of DITNS}
X-ray/radio-dim pulsar J0540-69 in LMC is one of the youngest and
most luminous rotation powered pulsars $^{14}$. It was first
observed as a 50 ms X-ray pulsar $^{15}$ and later it was
detected as a faint radio pulsar with a 640 MHz flux $\sim$0.4 mJy
$^{16}$. It's \.{P} value is 4.8$\times$10$^{-13}$ s/s
$^{14}$.

The average braking index of this pulsar is small (n$\sim$2.2 $^{17}$)
that it comes to a position close to DITNS RX
J0720.4-3125 in $\sim$10$^6$ yr which must be about the age of DITNS
RX J0720.4-3125 ($\sim$cooling age). Because of this,
J0540-69 can be in a former phase of at least some of the DITNSs.

Similar to J1846-0258, L$_X$/\.{E} of J0540-69 is large compared to
radio pulsars $^6$. When J0540-69 reaches to a position
close to the present position of RX J0720.4-3125 (which has
L$_X$/\.{E} $\sim$ 10), it will have L$_X$/\.{E} $>$ 1.

There is a pre-SN ring around J0540-69 similar to the rings around
SN1987A and the hourglass nebula around the blue supergiant Sher 25
$^{18}$. The pulsar is connected to oxygen-rich SNR 0540-6944
$^{19}$. There is a CO cloud located to the west of the
remnant but it is not clear if the SNR is interacting with it or not
$^{19}$. The region surrounding the SNR has a rich HI
structure $^{19}$. The pre-SN ring around this pulsar similar
to the rings seen around SN1987A $^{18}$ and its dense
environment suggest that the progenitor of this pulsar must be an
O-type star.

\subsection{X-ray pulsar J1811-1925; large braking index}
X-ray pulsar J1811-1925 is in a very dense medium $^3$.
The swept-up mass is 3-4 M$_{\odot}$ $^{20,21}$ and the
ejected mass should be considerably larger than the swept-up mass
$^{22}$. SNR G11.2-0.3 is similar to SNR Cas A in this sense
and the progenitor of it must be an O-type star (possibly early
O-type).

The spin down age of pulsars can be calculated as
\begin{equation}
t = \frac{P}{(n - 1) \dot{P}} [1 - (\frac{P_0}{P})^{n - 1}]
\end{equation}
where P$_0$ is the initial spin period of pulsar. SNR G11.2-0.3 is
probably related to the supernova (SN) of 386AD $^{2,23}$. 
If J1811-1925 is the compact remnant of SN 386AD, then its age (t) is 1618
years. Then, if its P$_0$$\sim$62 ms $^{2,24}$, its n must be $\sim$11.
If P$_0$ is less than 62 ms, then n must be larger than 11. For n=3 and 
P$_0$=62 ms, t is about 1970
years. If n=3 and t=1618 years, then P$_0$ must be 0.064666999929 s (i.e.
very close to the present P value of J1811-1925, which is 0.06466700 s).
This is not possible because the present \.{P} value of J1811-1925 is
4.40E-14 s/s and in 1618 years there must be a change in the P value at
least in the third digit after the decimal point unless n is much less
than 2. So, the realistic values are: t=1618 yr, P$_0$$\sim$62 ms, 
n$\sim$11. With 
such a braking index, J1811-1925 can reach to its present position
(P=0.06466700 s and \.{P}=4.40E-14 s/s) on the P-\.{P} diagram in
1618 years starting with P$_0$$\sim$62 ms.

So, this pulsar seems to be evolving downwards on the
P-\.{P} diagram with a sharp decrease in \.{P} and very small
increase in P. Radio/X-ray pulsars in the region $\tau$ = 10$^5$ -
2$\times$10$^7$ yr are displayed in Figure 1.
Most of such pulsars are located on the lower left part of
the diagram. On this part of the diagram, there are also some radio
pulsars which have not been detected in X-rays. X-ray pulsar
J1811-1925 seem to be evolving in the direction to the region of
these radio/X-ray and radio pulsars. X-ray luminosity of
J1811-1925 is $\sim$10$^{33}$-10$^{34}$ erg/s and the X-ray
luminosity of the radio/X-ray pulsars in this region is
10$^{30}$-10$^{34}$ erg/s $^{25,26}$.

The large n value of J1811-1925 indicates that either there is magnetic 
field decay (see Geppert \& Rheinhardt $^{27}$ for the magnetic field 
decay in neutron stars and see Bisnovatyi-Kogan $^{28}$ for a review on 
magnetic field of neutron stars in general) or the angle between the spin 
axis and the magnetic field axis is decreasing sharply. 

The lack of detected radio emission from J1811-1925 may be due to
selection effects (luminosity function or beaming fraction) or maybe
the radio emission of this X-ray pulsar is screened by some dense
plasma around it.

AXP 1E 2259+586 is connected to SNR CTB 109 and the age of this SNR
is $\sim$10$^4$ yr $^{29-31}$ that AXP 1E 2259+586
has n$\sim$40. Note that J1811-1925 must have such a very large n if
it has P$_0$$\sim$10 ms which is generally assumed to be a typical
P$_0$ value for radio pulsars. Although they may have similar slope
of tracks on the P-\.{P} diagram, note that AXP 1E 2259+586 and
J1811-1925 are located on very different parts of the P-\.{P}
diagram (see Fig.1) that their origins must be different.

\subsection{A comparison between J1846-0258, J1811-1925, and
J0540-69}

All these 3 pulsars must have O-type progenitors. The progenitors of
J1811-1925 and J0540-69 can be more massive than the progenitor of
J1846-0258, and possibly the progenitor of J1811-1925 can be more
massive than the progenitor of J0540-69. Since their positions on
the diagram and their braking indices seem to be very different
compared to each other, different physical processes must be
dominant for each of them. Mass of the neutron star can be the main
parameter for different physical processes to occur. Other than the
mass of the progenitor, closeness of the binary and mass of the
companion star before the SN explosion (i.e. before the disruption of the 
binary system), and the fact that the progenitor might be isolated
must be considered to understand the formation of different types of
neutron star.

If the progenitors of these 3 pulsars had actually very different
zero age main sequence mass values compared to each other, these
pulsars may have very different masses compared to each other.
Differences in their n and \.{P} values, and their positions on the
diagram together with the observational data about their
environments support this idea. Among these 3 pulsars, J1846-0258
has the largest \.{P} (comparable to some AXPs) and the smallest n
value that it may have the smallest mass (possibly $\sim$0.5-0.7
M$_{\odot}$). Similarly, J0540-69 may be less massive than J1811-1925.

X-ray pulsars J1846-0258 and J1811-1925 are two extreme cases and their 
initial parameters must be very different compared to each other. 
Differences in their mass and initial magnetic field (or 
\.{P}$_{initial}$) values must be the main factors for the significant 
difference between their evolutions. These differences must be the cause 
of the difference between their n values; in the case of J1846-0258, the 
neutron star may have smaller mass and larger radius (extended atmosphere) 
and higher initial magnetic field so that it has large (and possibly 
increasing) \.{P}, whereas in the case of J1811-1925, the neutron star 
must have lower initial magnetic field and possibly much more massive than 
J1846-0258 that its B value drops sharply due to either magnetic field 
decay or a decrease in the angle between the magnetic field and the spin 
axes.   

In order to understand if some of these pulsars evolved through
massive binaries, central regions of the SNRs connected to these 3
pulsars must be examined to see if there are O or B-type stars
present which are connected to the SNRs. Failing to find such OB
stars may lead to the fact that the progenitors were either isolated
or in binary systems with later type companions, but also note that
the sources being located at large distances can make it difficult
to realize the presence of such OB stars.

\section{Conclusions}

X-ray pulsar J1846-0258 seems to be evolving to become SGR/AXP in
$\sim$10$^4$ yr. It can be a low-mass neutron star which has an
O-type progenitor.

X-ray/radio pulsar J0540-69 can become a DITNS in $\sim$10$^6$ yr.
Its progenitor must also be O-type, possibly more massive than the
progenitor of J1846-0258.

X-ray pulsar J1811-1925 has a small \.{P} which is decreasing
sharply. This X-ray pulsar can be in a former phase of the radio/X-ray
and radio pulsars on the lower left part of the P-\.{P}
diagram (Fig.1). The progenitor of J1811-1925 must be an O-type
star, possibly more massive than the progenitors of the other 2
pulsars.

Which type a neutron star will become must depend on its
progenitor's mass and on the effects of the binary evolution. For
the 3 pulsars examined in this work, mass of the progenitor may be
the main parameter for the mass and hence the type of the neutron
star.

{\it Acknowledgments}
This work is supported by Bo\u{g}azi\c{c}i University and Middle East 
Technical University.

\clearpage

{\bf Figure Caption} \\
{\bf Figure 1:}
Spin period vs. time derivative of spin period diagram of
different types of pulsar. Filled circles represent Galactic X-ray
pulsars J1846-0258 and J1811-1925, and X-ray/radio pulsar J0540-69
in LMC. Filled squares and open squares display SGRs
and AXPs, respectively. Asterisks represent DITNSs
(for 3 of them the upper limits on
their \.{P} values are shown with arbitrary arrows). DRQNSs
are displayed with sign '+'. Radio pulsars are shown
as small dots. Radio/X-ray pulsars in the region $\tau$ = 10$^5$ -
2$\times$10$^7$ yr are shown with sign 'X'. B12-B14, E32-E38, and
A3-A7 indicate B=10$^{12}$-10$^{14}$ G, \.{E}=10$^{32}$-10$^{38}$
erg/s, and $\tau$=10$^3$-10$^7$ yr, respectively. The P and \.{P}
values of DITNSs are from Haberl $^{32}$ and the P and \.{P} values
of radio/X-ray pulsars are from Becker \& Aschenbach $^{25}$ and 
Possenti et al. $^{26}$. All the other P and \.{P} values are from 
ATNF catalogue $^1$.

\clearpage

\begin{figure}[t]
\vspace{3cm}
\includegraphics{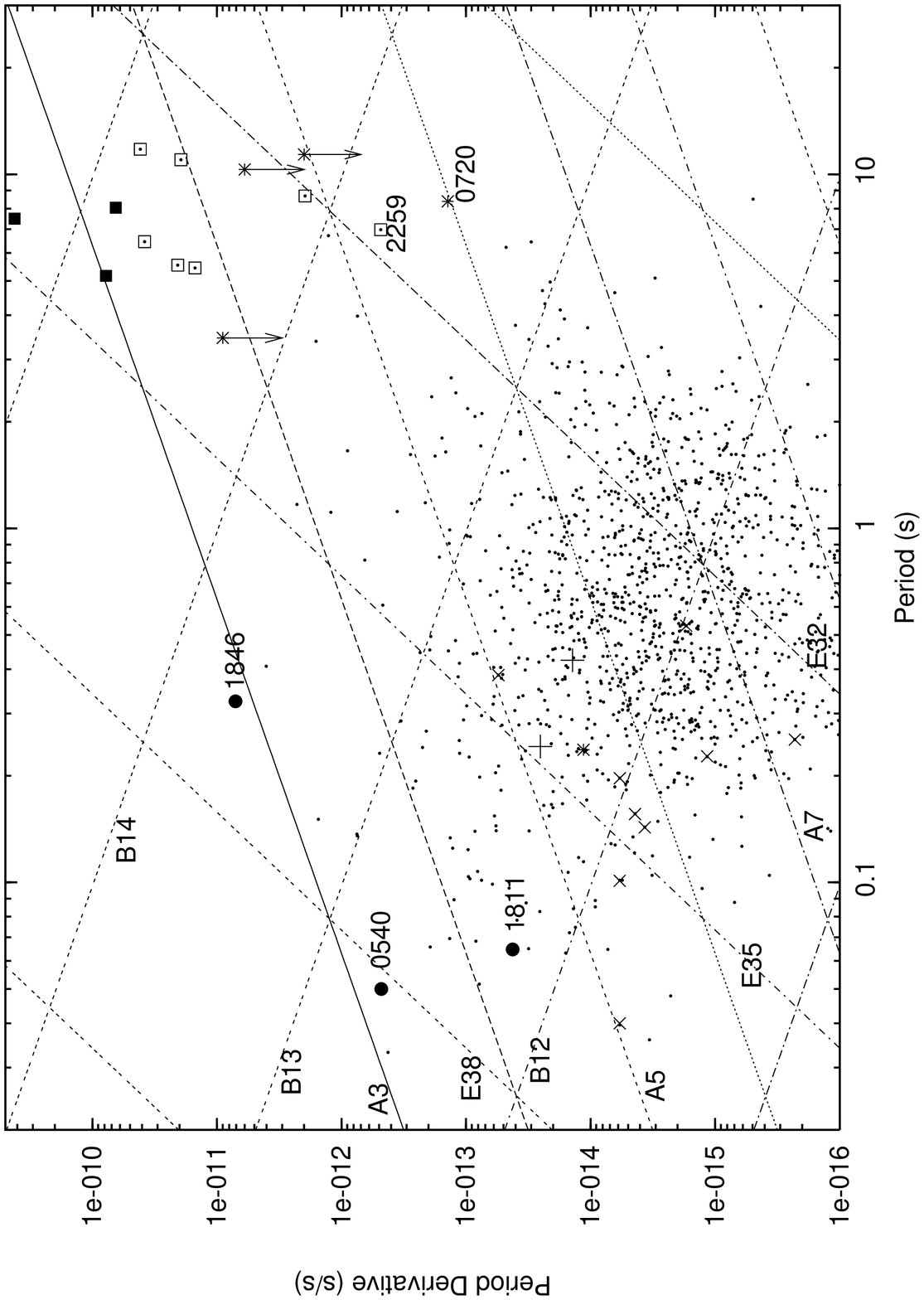}
\end{figure}

\end{document}